\documentclass[authoryear,preprint,review,12pt]{elsarticle}
\usepackage[
  letterpaper,
  margin=1in
]{geometry}

\usepackage{amssymb}
\usepackage{amsmath}
\usepackage{pifont}
\usepackage{float}
\usepackage{nameref}
\usepackage{url}

\usepackage[x11names]{xcolor}
\usepackage[
  colorlinks,
]{hyperref}

\journal{arXiv}

\begin{document}

\begin{frontmatter}

\title{Lakeplace: Sensing interactions between lakes and human activities} 


\author{Meicheng Xiong}
\author{Di Zhu\corref{cor1}}
\cortext[cor1]{Corresponding author}
\ead{dizhu@umn.edu}
\address{Department of Geography, Environment and Society, University of Minnesota, Twin Cities, Minneapolis, USA}

\begin{abstract}
Urban freshwater ecosystems, composed of rivers, ponds, lakes, and other water bodies, have essential socioeconomic and ecological values for urban residents. However, research investigating how individuals interact with lakes remains limited, especially within cities and at fine spatiotemporal resolutions. To fill this gap, we propose a data-driven analytical framework that comprehensively senses human-lake interactions and profiles the social-demographic characteristics of intra-city lakes. The term ``lakeplace" is proposed to depict a place containing lakes and human activities within it. For each lake, the geographic boundary of its lakeplace refers to the first-order administrative units, reflecting the neighboring scale of lake socioeconomics. Utilizing large-scale individual mobile positioning data, we performed lakeplace sensing on the 2,036 major lakes in the Twin Cities Metropolitan Area (TCMA), Minnesota, and the people interacting with them. The popularity of each lakeplace was measured by its temporal visitations and further categorized as on-lake and around-lake human activities. Popular lakeplaces were investigated to depict whether the attractiveness of a lake is mostly brought by the lake itself, or the social-demographic environment around it. The lakeplace sensing framework offers a practical approach to the spatiotemporal characteristics of human activities and understanding the social-demographic knowledge related to human-lake systems. Our work exemplifies the social sensing of human-environment interactions via geospatial big data, shedding light on human-oriented sustainable urban planning and urban water resource management.
\end{abstract}

\end{frontmatter}




\section{Introduction: From lakes to lakeplaces}
\label{sec1}
Areas covered by water have long been the birthplaces of civilizations, furnishing resources as well as trades for generations \citep{yasuda2012water, hosseiny2021water}. Being preferred sites of human settlement, places near seas, rivers, and lakes are where humans interact with water intensively \citep{riera2001nature}. Globally, over 117 million lakes are scattered in urban and rural areas \citep{verpoorter2014global}. Lakes, varying in their size and function, form diverse freshwater landscapes and ecosystems that serve as living environments for residents, and become human's lakes shaped by both natural processes and humanistic influences.

Understanding a human's lake can be straightforward. The paradigm making the human's lake begins by attaching human experience to a lake, which embraces the notion of Yi-fu Tuan's humanistic place:``Location is transformed into place when it is established as significant'' \citep{tuan2017humanistic}. The essence of place appears to be the geography of human significance. In this context, a human's lake is seen as a lake saturated by human-lake interactions. A human's lake can be either private or public, corresponding to individual experiences and collective moments, respectively. The existence of a lake is acknowledged directly via the sense of lakes, involving one or multiple individuals' interactions with, feelings about, and expressions of the lakes.

There are no remarkable obstacles to understanding a human's lake from the individual perspective, as numerous studies narrated the story of lakes in the eyes of different individuals \citep{cantrill1998environmental,simoni2015exploration,ebner2022combining}. It is much more challenging, however, to profile a human's lake through the perspective of multiple individuals. Theoretically, it can be accomplished by conducting surveys and interviews with several individuals and groups \citep{stedman2007perceived, mitroi2022urban}, in which way a subjective human's lake can be captured through the gathering of collective perception. The gap is: very little is known about the spatiotemporal characteristics of collective human activities, especially at a fine scale. 
To address the gap, geospatial big data, which traces the whereabouts of numerous individuals, can capture detailed human activities by offering rich and high-resolution spatiotemporal information \citep{li2016geospatial,shaw2021understanding}. The application of geospatial big data also helps to mitigate the traditionally high costs of human activity data collection \citep{huang2021analytics}. Thus, it enables the sense of not only one single or several, but a larger number of human's lakes scattered in the urban water system.

\begin{figure}[H]
  \centering
  \includegraphics[width=0.6\paperwidth]{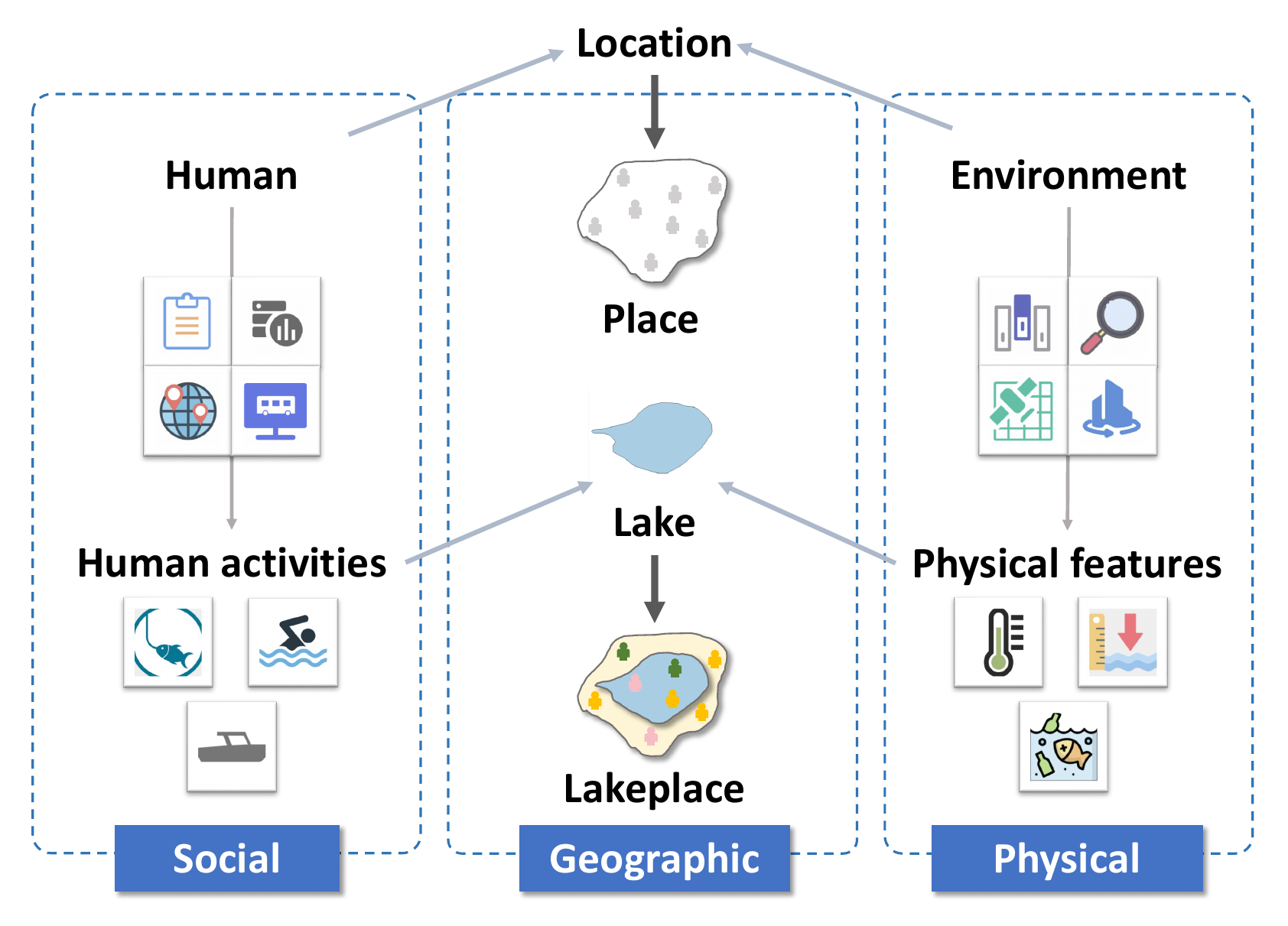}
  \caption{From lakes to lakeplaces: lakes enriched with human-lake interactions}
  \label{fig:01}
\end{figure}

Lake-related human activities characterized via geospatial big data imply a promising path to sensing human's lake, as illustrated in Figure \ref{fig:01}. Adopting Yi-fu Tuan's notion of place, we propose to go from lakes to lakeplaces, coalescing a data-driven knowledge discovery on human-lake interactions. A lake, with its locational features, is first put ahead of two dimensions, namely human and environment, to form an abstract place. The environmental dimension of a lake can be depicted by an array of physical features such as temperature, sediments, and pollutants. More importantly, the human dimension is characterized by lake-related activities like on-lake fishing, swimming, and boating, as well as around-lake jogging, recreation, and social demographics such as income, race, and age of visitors. Attaching environment and human dimensions to the location reflects the practice that identifying lakes based on sensible physical and social features, which is traditionally achieved by remote sensing images, map services, survey data, etc. Once the human's lake is ensured, a lake is turned into \textit{lakeplace}, referring to the place where people interact with the lake. Distinct from similar terms such as lake district \citep{cooper2011mapping}, lake region \citep{ellis1997sustainable}, and lake area \citep{ji2005palaeoclimatic}, \textit{lakeplace} centers human experience attached to the lakes, beyond the location and physical attributes of places. More generally, the relationship between humans and other objects sharing similar physical traits to lakes, such as a park, can be investigated similarly.

Practically, the geographic extent of a lakeplace refers to the area covering and surrounding one or multiple lakes at a certain spatial scale, such as census block, census block group, and census tract. In this study, we generated the lakeplace layer based on census block and developed a data-driven lakeplace sensing framework (Section \ref{framework}) to sense and profile urban lakes based on fine-scale human-lake interactions. Utilizing geospatial big data, each lakeplace is enriched with spatiotemporal features of human activities and mobilities, combined with the demographic characteristics of lake visitors.

\section{Related works: Sense of lakeplace}
Sensing lakeplace can be approached as a practice of Sense of Place (SOP) \citep{shamai1991sense}, which is a cover-all concept and can be interpreted as human attachments to the place: a place is established as a meaningful location of human experience \citep{jorgensen2011measuring}. We have witnessed variants inheriting and developing this human-oriented idea, such as ``Places as locations of human significance'' \citep{cresswell2008place}, ``Places are locations with meaning'' \citep{cheshmehzangi2014spatial}, and ``Places only become significant when they are given meaning'' \citep{thomasson2019geographic}, laying the theoretical ground for the sense of place. To sense the place is to discover and examine interactions between the human and the environment. It allows for advanced understandings of places, shedding insight on place portrayal and management \citep{davenport2005getting}.

A plethora of studies have examined all sorts of places by investigating the relationship between humans and different types of places. Early studies mostly discussed the lifetime human experience and emotions attached to the dwelling space, where people spend most time \citep{hay1998sense}. Typically, the place in such a context is a certain administrative unit or the area near the known landscapes \citep{cuba1993constructing}. Areas near seas, rivers, and lakes, as popular settlements, have also been intensively examined \citep{jorgensen2001sense, baindur2014bangalore}. The sense of place has also covered other significant places in human daily life, such as the workplace and recreation space. Farms \citep{mullendore2015us}, schools \citep{cumming2015australian}, urban green open space \citep{vzlender2020testing}, healthcare space \citep{andrews2004re}, and parks \citep{mehnen2013governance} are all places of interest, reflecting various venues of human-environment interactions. Literature on water bodies functioning goes beyond the residential environment, involving studies of water landscapes in recreational spaces as experienced by visitors \citep{wartmann2018investigating}. In light of this, water bodies as both residential and non-residential places imply multi-functions of urban freshwater ecosystems as well as human-water interaction diversity. Such diversity serves as the foundation for the multifaceted and enriched sense of lakeplace.

The recent boom of crowd-sourcing techniques and geographic sciences has marked a new era for the sense of place, with the wide application of big spatiotemporal data and social sensing in urban scenarios \citep{liu2015social}. Compared with traditional data represented by surveys and interviews utilized in the aforementioned studies, big spatiotemporal data such as social media check-ins, Google Street Views, and GPS data captures semantic information of human activities at finer spatial and temporal scales \citep{deng2019geospatial, yang2024street}. This enables researchers to better understand the complex interactions between humans and the environment. Place-oriented studies can be roughly divided into two categories along this line. One is mainly illustrating and tagging places with human activities and perceptions, such as uncovering place types \citep{zhu2020understanding}, landscape elements \citep{zhao2023sensing}, and urban dynamics \citep{silva2024actionable}. Another one is more task-driven, covering studies in detecting places where social events or natural disasters happen \citep{zhou2020understanding,spruce2021social}. Both categories can be found in data-driven studies exploring the human-water interactions, which demonstrate the sense of urban freshwater ecosystems in regular and urgent circumstances, respectively \citep{dong2022predictive}.

Despite these efforts, most existing work concerning urban freshwater ecosystems has focused on the impact of human-water interactions on water quality, sediments, or elements. Our study, instead of digging into how human activities lead to biophysical transformation of lakes, is anchored on the exploration of portraying the humanistic facet of lakes \citep{zhao2022humanistic}.

\section{Twin Cities: a populous land of thousand lakes}
\begin{figure}[H]
        \centering
        \setlength{\abovecaptionskip}{0.cm}
        \includegraphics[width=0.8\textwidth]{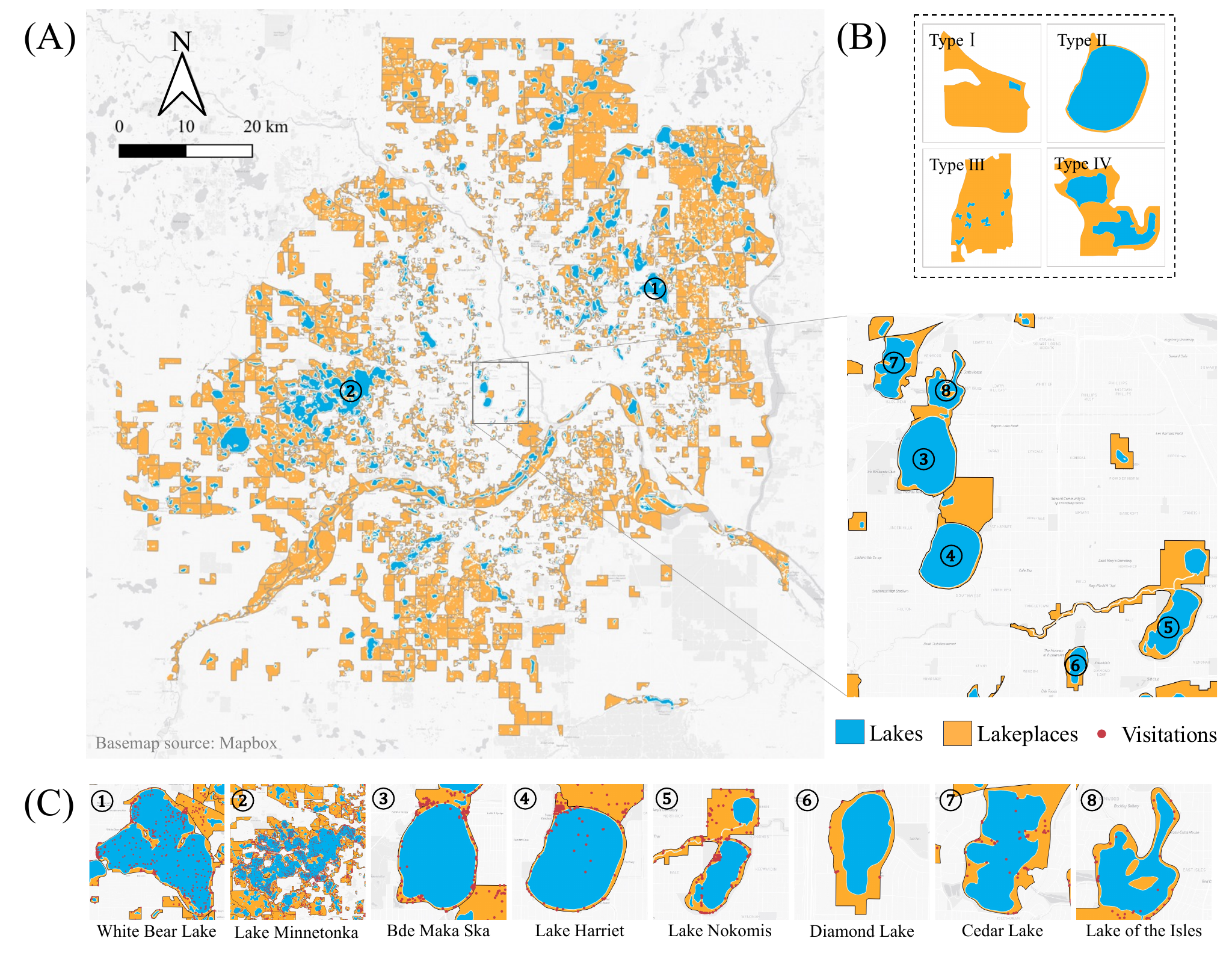}
        \caption{A land of thousand lakes: the Twin Cities Metropolitan Area. (A) presents all the lakes and lakeplaces inside the TCMA, along with the zoomed-in view to show the downtown and uptown area. \uppercase\expandafter{\romannumeral1}) to \uppercase\expandafter{\romannumeral4}) in (B) show four types of spatial relationships between lakes and lakeplaces. \ding{172} to \ding{179} in (C) present human visitations inside and around eight popular lakes, namely White Bear Lake, Lake Minnetonka, Bde Maka Ska, Lake Harriet, Lake Nokomis, Diamond Lake, Cedar Lake, and Lake of the Isles. For simplicity, north arrows and scales are omitted in the following figures.}
        \label{fig:02}
        \vspace{-0.2cm}
    \end{figure}

The Twin Cities Metropolitan Area (TCMA), situated in the core area of Minnesota, is commonly known as the ``Land of 10,000 Lakes''. Covering the rural, suburban, and urban areas of Minneapolis–Saint Paul and accounting for more than half of Minnesota’s total residents, TCMA is well endowed with thousands of natural and artificial lakes, rendering TCMA a land mixed of natural and humanistic features \citep{wright1989origin}. Of all the lakes inside the TCMA, over 500 lakes are larger than ten hectares \citep{kloiber2002procedure}, such as White Bear Lake (Figure \ref{fig:02}\ding{172}), Lake Minnetonka (Figure \ref{fig:02}\ding{173}), Bde Maka Ska (Figure \ref{fig:02}\ding{174}), Lake Harriet (Figure \ref{fig:02}\ding{175}), Lake Nokomis (Figure \ref{fig:02}\ding{176}), Diamond Lake (Figure \ref{fig:02}\ding{177}), Cedar Lake (Figure \ref{fig:02}\ding{178}), and Lake of the Isles (Figure \ref{fig:02}\ding{179}). More unnamed lakes, along with these well-known lakes, form a large urban freshwater ecosystem serving urban residents. They not only construct the land use and urban structure of the Twin Cities, but also play a pivotal role in shaping human activities as embedded spaces in the multi-functional areas. By providing multiple functions such as recreation, shipping, and education for people, lakes interact with Twin Cities' residents and visitors, and vice versa.

Adopting the lakeplace sensing framework (Section \ref{framework}), we obtained 1,738 lakeplaces from water bodies in the TCMA, each of which contains one or multiple lakes. Four basic types of lake-lakeplace spatial relationship are shown in the subfigures \uppercase\expandafter{\romannumeral1}) to \uppercase\expandafter{\romannumeral4}) of Figure \ref{fig:02}B, depicted by two factors, namely match type and water area ratio, respectively. As for the match type, there may be one single lake (Type \uppercase\expandafter{\romannumeral1} and Type \uppercase\expandafter{\romannumeral2}) or several lakes inside one lakeplace (Type \uppercase\expandafter{\romannumeral3} and Type \uppercase\expandafter{\romannumeral4}). For a situation where only one lake is inside a lakeplace, the lake may be a small and isolated pond located in a residential or commercial area (e.g. a tiny and nameless pond, Type \uppercase\expandafter{\romannumeral1} in Figure \ref{fig:02}B), or a famous and popular lake (e.g. Lake Harriet, Type \uppercase\expandafter{\romannumeral2} in Figure \ref{fig:02}B). While for the other cases where more than one lakes are located in the same lakeplace, lakes usually impress people as a cluster of small ponds (e.g. a series of small ponds, Type \uppercase\expandafter{\romannumeral3} in Figure \ref{fig:02}B), or several discernible lakes which are close to each other (e.g. Lac Lavon and Keller Lake, Type \uppercase\expandafter{\romannumeral4} in Figure \ref{fig:02}B). Types \uppercase\expandafter{\romannumeral3} and \uppercase\expandafter{\romannumeral4} are usually observed in nature-oriented recreation districts such as lake parks and golf course fields, or suburban areas.
    
We employed a large-scale mobile phone positioning dataset provided by PlaceIQ that covers approximately 5\% of the U.S. population and has been widely used in studies of human mobility \citep{weill2020social, couture2022jue}. The original data records users' locations and time stamps whenever they open their mobile devices. Each record contains a unique device ID, time, coordinate (in longitude-latitude format), and stay duration of the user, supporting the tracing of individual movements and their spatiotemporal patterns. We chose the month of July 2021 as the study period. The amount of data is sufficient for capturing human activities, as over 720,000 individuals and about 21,000,000 records were observed, accounting for around 20\% population in the TCMA.
    
We applied multiple data sources to support the lakeplace construction and exploration, including the block-level zoning layer, waterbody layer, and POI data. The block-level zoning layer of TCMA, provided by the National Historical Geographic Information System (NHGIS) website (\url{https://www.nhgis.org/}), is utilized to generate lakeplaces. The water body boundary layer for the year 2021 was obtained from OpenStreetMap (\url{https://www.openstreetmap.org/}). In this study, we started with all 18,964 lakes inside the TCMA, and then selected 2,906 major lakes that are larger than one hectare based on the criteria of the LAGOS dataset \citep{cheruvelil2021lagos} to ensure that each lakeplace contains at least one lake. As a result, 9,327 lakes contained in lakeplaces were investigated. POI data is provided by SafeGraph, with 43,193 records covering various socioeconomic categories inside the TCMA. We reclassified POIs into 11 categories, including travel, retail, dining, entertainment, automotive, business, real estate, financial, educational, industrial, and transportation.

\section{Results: Evidence from lakeplace sensing}
\begin{figure*}[htbp]
        \centering
        \setlength{\abovecaptionskip}{0.cm}
        \includegraphics[width=0.7\textwidth]{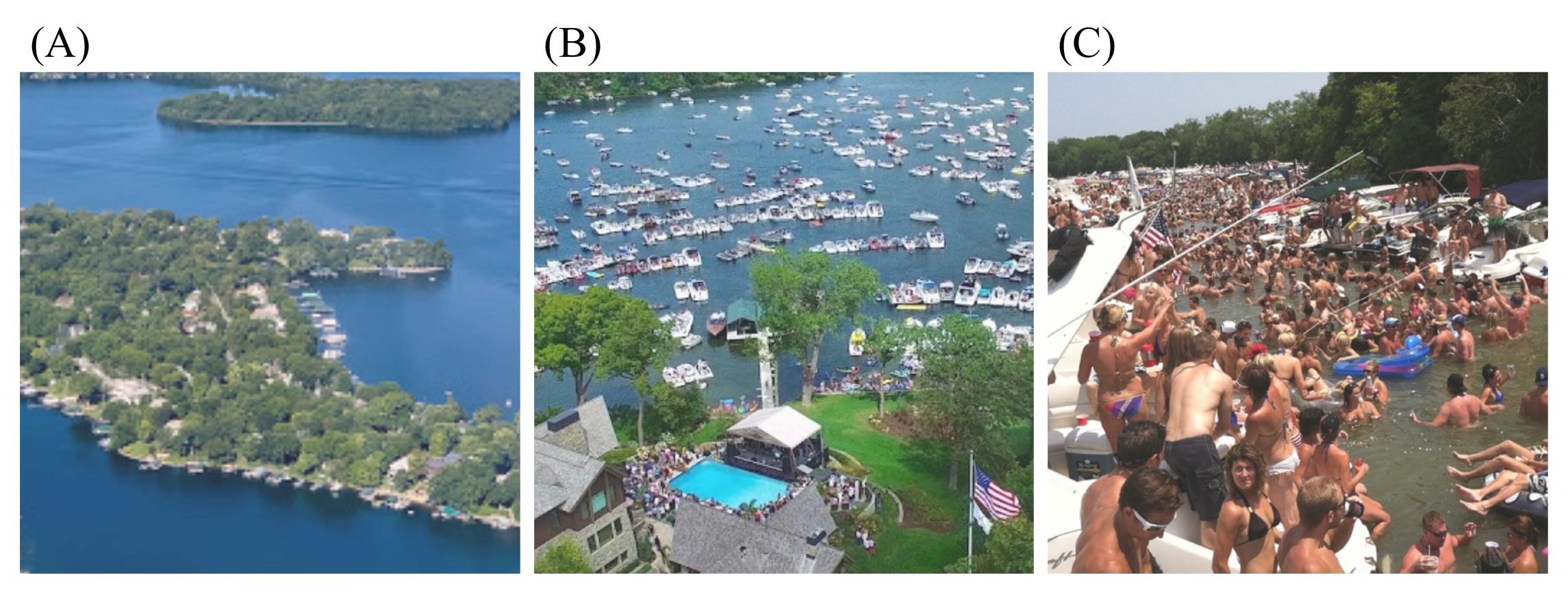}
        \caption{Three photos of Lake Minnetonka in the TCMA. (A) shows the lake without people, (B) presents the zoom-out view of people on and around the lake, and (C) exhibits the zoom-in view to show the crowd interacting with the lake.}
        \label{fig:03}
        \vspace{-0.2cm}
    \end{figure*}

Lakeplace sensing answers three research questions: \textit{which lakeplaces are popular}, \textit{why they are popular}, and \textit{who interact with them}? That said, we consider lakeplace popularity to be a central theme that encapsulates multiple facets of human–lake interaction. Figure \ref{fig:03} displays three photos of the Lake Minnetonka area (Figure \ref{fig:02}\ding{173}) captured at different times, telling different stories regarding the popularity of the same lakeplace. Figure \ref{fig:03}A presents the lake view without people, delivering no sign of on-lake human activities. There are nonetheless plenty of people in Figure \ref{fig:03}B, which indicates another facet of the lakeplace, which is a place full of human dynamics. When zooming in on the crowd (Figure \ref{fig:03}C), diversity is uncovered as heterogeneity raises more concerns about the social demographics of these people. The three questions above can be linked as a following stream: from lakeplace without people to lakeplace with people, and to lakeplace with different people. It was concretized by the following three sections of our case study.
    
\subsection{Visitation: time and space}
\begin{figure}[H]
        \centering
        \setlength{\abovecaptionskip}{0.cm}
        \includegraphics[width=0.8\textwidth]{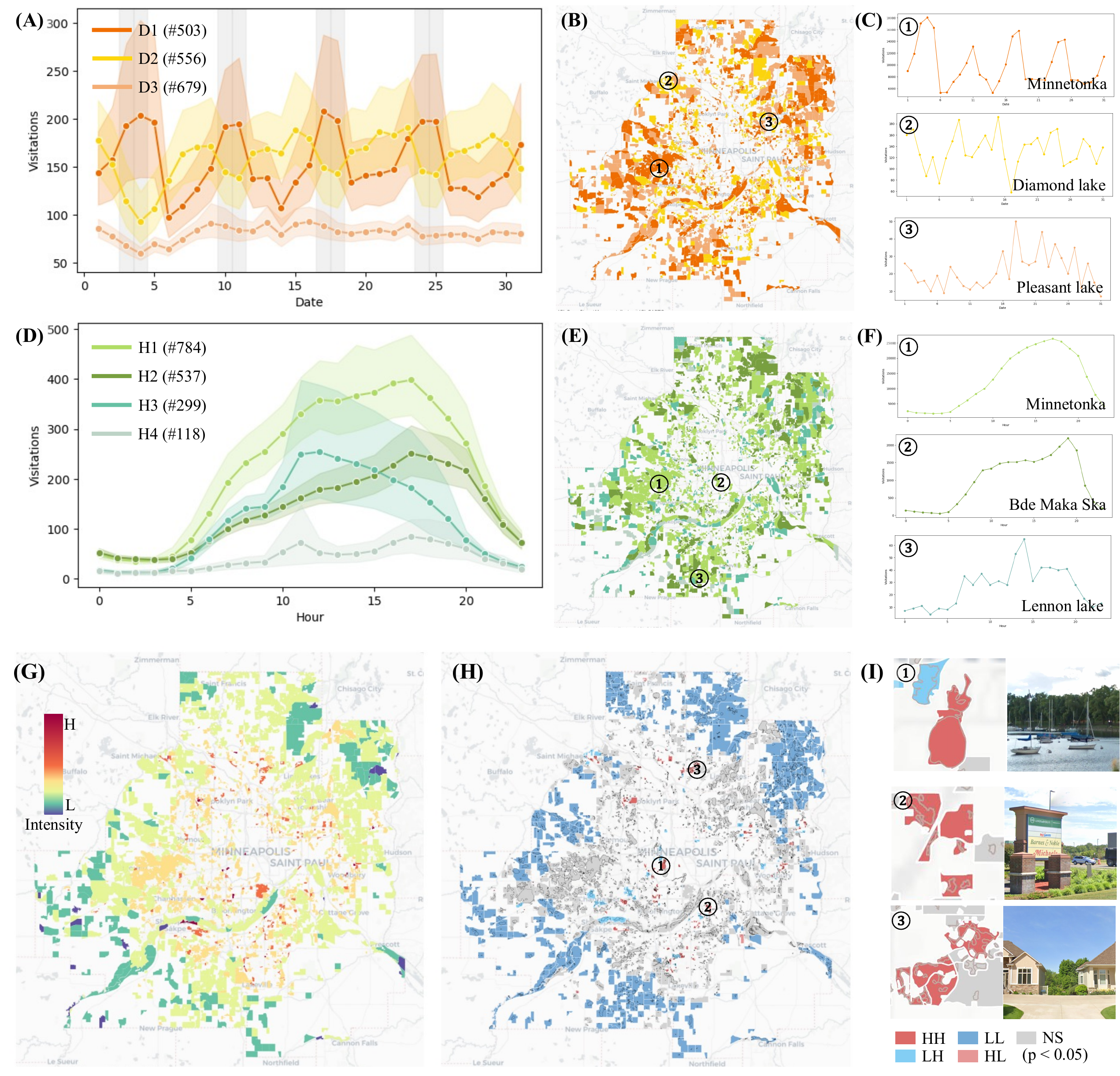}
        \caption{(A) to (F) present temporal pattern of daily visitations and hourly visitations. (A) and (D) present the daily and hourly visitation variations inside the lakeplaces of different clusters, respectively. Note that the number behind D or H refers to the number of lakeplaces belonging to this type, and the gray areas in (A) refer to the weekends. (B) and (E) show the spatial distribution of lakeplaces labeled by daily cluster and hourly clusters, along with typical lakeplaces belonging to different clusters in (C) and (F), respectively. (G) to (I) show the spatial pattern of daily visitations. (G) presents intensity calculated by lakeplace daily visitations. (H) shows results of local Moran's I and three examples of intensity hotspots, each containing several lakeplaces. The zoomed-in views along with Google Street Views of three examples in (I) indicate that three sets of lakeplaces are located in the recreational, commercial, and residential areas.}
        \label{fig:04}
    \end{figure}

Visitations inside the TCMA in July 2021 are utilized to describe the temporal and spatial variations of lakeplaces. We extracted the hourly average visitations and daily average visitations for each lakeplace. To capture the temporal variation characteristics, the agglomerative clustering method was applied to cluster lakeplaces based on hourly and daily variation signatures, respectively. As a result, lakeplaces can be classified into three types based on their daily variations and four types based on their hourly and diurnal variations, respectively. From Figures \ref{fig:04}A-C, one can easily observe that D1 donates lakeplaces with higher visitations on weekends, such as Lake Minnetonka (Figure \ref{fig:02}\ding{173}). Lakeplaces belonging to D2 show an opposite pattern to those of D1, as they are more frequently visited during weekdays than weekends. Compared to D1 and D2, D3 represents those ``unpopular lakeplaces'': they are rarely visited no matter on weekdays or weekends. Similarly, H1, H2, H3, and H4 represent the four types of lakeplaces based on their hourly variations (Figures \ref{fig:04}D-F): high during the whole day (H1), higher in the afternoon (H2), higher in the morning (H3), or low during the whole day (H4). Visitation disparity between weekdays and weekends, to some extent, refers to the difference between working and recreational functions, while the variation during the day is naturally connected to the ``residential or work'' dichotomy.

Similarly, we employed daily average visitation to uncover spatial patterns. To mitigate the size bias, we obtained the visitation intensity for each lakeplace by dividing the total visitations by its area size. Figure \ref{fig:04}G shows that most lakeplaces with relatively high intensity are scattered in the interior of the TCMA, while those with low intensity are clustered in the periphery of the TCMA. The spatial autocorrelated pattern is also proved by the significant positive value of global Moran's I (0.24). Moreover, we identified several significant clusters and outliers based on local Moran's I (Figure \ref{fig:04}H). Coldspots (LL) account for the largest proportion of significant local patterns, most of which are located in the periphery of the TCMA. Hotspots (HH) can be observed inside the TCMA, where several lakeplaces inside the downtown area (\ding{172} in Figure \ref{fig:04}H), the commercial area near the Twin Cities Premium Outlets shopping center (\ding{173} in Figure \ref{fig:04}H), and the residential area in Blaine (\ding{174} in Figure \ref{fig:04}H) are identified as examples. The result indicates that popular places are still significantly hot from the perspective of lakeplace.
    
\subsection{Popularity: which and why?}
Human activities such as swimming, fishing, and boating are typical on-lake human activities, indicating more intense human-lake interaction. To measure popularity quantitatively, we calculated and ranked human activity intensity $\alpha_t$ for each lakeplace (Section \ref{framework}, Equation \ref{Equation 1}). It reflects human activity density brought by on-lake and other human activities inside the lakeplace. To explore the reasons behind popularity, we employed the lake contribution $c_t$ to represent the on-lake human activity proportion of all human activities (Section \ref{framework}, Equation \ref{Equation 2}). It illustrates how humans interact with the lakes.

\subsubsection{On-lake contributions}\label{sec:On-lake}
\begin{figure}[H]
        \centering
        \setlength{\abovecaptionskip}{0.cm}
        \includegraphics[width=0.8\textwidth]{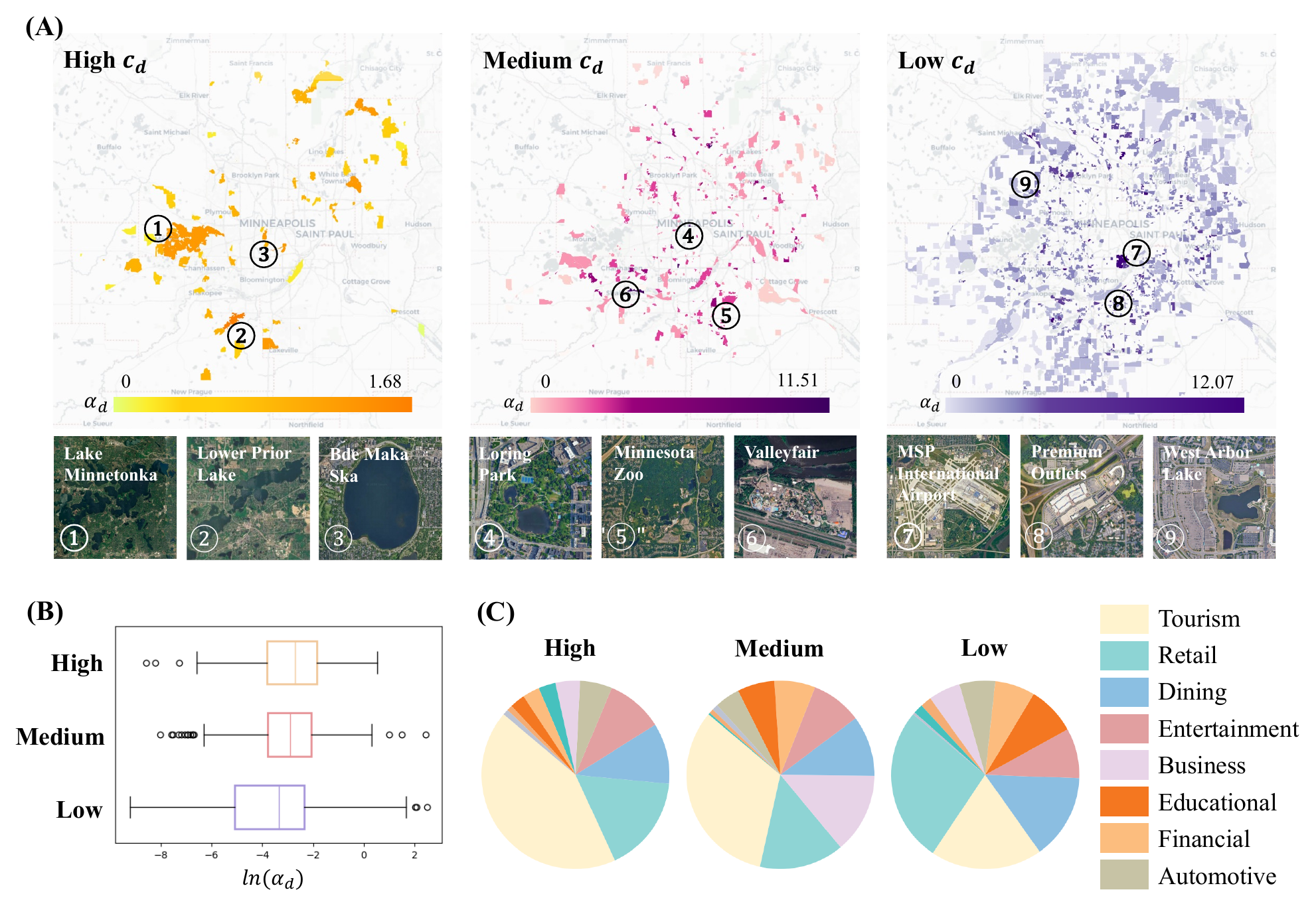}
        \caption{Three types of lakeplaces based on on-lake contribution levels: high (4.9\%), medium (20.6\%), and low (74.5\%). The top representative popular lakeplaces within each category are displayed in the remote sensing images in (A), with the more popular lakeplaces of each lake-contribution level in darker colors. The range of $\alpha_d$ for all lakeplaces of three types is presented in (B), and their POI composition is shown in (C).}
        \label{fig:06}
    \end{figure}

We first investigated daily lakeplace popularity by employing daily human activity intensity $\alpha_d$, where $d$ means the temporal scale is set to be daily. Accordingly, daily on-lake contribution $c_d$ was calculated based on the daily on-lake activity proportion to all activities. Given that, we hereby divide lakeplaces into three types: low on-lake contribution ($c_d < 0.01$), medium on-lake contribution ($0.01 \leq c_d < 0.1$), and high on-lake contribution ($0.1 \leq c_d < 1$). As shown in Figure \ref{fig:06}, a significant disparity exists among human activity intensity and land use composition of each type. Lakeplaces with high on-lake contribution (4.9\% of all lakeplaces) are where large-size and well-known lakes are located, attractive to visitors in or outside the TCMA. The visitors to these lakeplaces mainly aim to interact with lakes for on-lake activities or activities right around lakes, and the size proportion of lakes to such lakeplaces is relatively higher. The most popular lakeplaces of this type are popular lake sights in the TCMA, with Lake Minnetonka (\ding{172} in Figure \ref{fig:06}A), Lower Prior Lake (\ding{173} in Figure \ref{fig:06}A) and Bde Maka Ska (\ding{174} in Figure \ref{fig:06}A) ranked to be the top three popular ones. Most of the other lakeplaces with relatively high $\alpha_d$ are constructed and perceived for tourism, being perceived as the typical ``lakeplaces'' due to the predominant role of lakes.

When on-lake contribution decreases to a medium level (20.6\% of all lakeplaces), the functions of lakeplaces turn to be more daily recreation-oriented, rather than tourism. The most popular ones mainly contain lake parks near residential areas, amusement parks, and zoos, characterized by smaller lake sizes, fewer direct human-lake interactions but higher overall intensity. Loring Park (\ding{175} in Figure \ref{fig:06}A), Minnesota Zoo (\ding{176} in Figure \ref{fig:06}A) and Valleyfair (\ding{177} in Figure \ref{fig:06}A) are three typical ones. Compared with lakeplaces with high-level on-lake contributions, there are more facilities for business, education, and auto inside lakeplaces of this type (Figure \ref{fig:06}C). The phenomena reflect that with the decrease of human-lake interaction intensity, the lakeplaces become more daily life-oriented places, as they become more central to the routines of residents.

Most lakeplaces (74.5\% of all lakeplaces) have a low-level on-lake contribution. They tend to be at local centers near highways, which are usually filled with local services such as gas stations, stores, and restaurants. The visitors gather in these lakeplaces mainly for services provided by these facilities near lakes, instead of direct human-lake interactions. The most popular lakeplaces within this category include transportation and business centers, such as MSP International Airport (\ding{178} in Figure \ref{fig:06}A), Premium Outlets (\ding{179} in Figure \ref{fig:06}A) and West Arbor Lake (\ding{180} in Figure \ref{fig:06}A). It is worth noting that $\alpha_d$ for this low on-lake contribution type shows the largest variation, referring to the most popular lakeplaces as well as the most unpopular lakes. That said, these life-service places can be either packed or neglected.

By considering the on-lake contributions, we treated lakeplaces not only based on the overall human activity intensity. There is a possibility that two lakeplaces share a similarly high human activity intensity, but can be differentiated by their on-lake contributions, indicating their heterogeneous characteristics and functions. For example, Valleyfair ($\alpha_d=11.52$, \ding{177} in Figure \ref{fig:06}A) and Premium Outlets ($\alpha_d=12.07$, \ding{179} in Figure \ref{fig:06}A) are popular lakeplaces sharing a similar human activity intensity. Nevertheless, the former belongs to the medium-level on-lake contribution type ($c_d = 0.05$), while the latter is a typical one of the low-level on-lake contribution type ($c_d = 0.003$).

\subsubsection{Comparing weekdays and weekends}\label{sec:Comparing}
\begin{figure*}[htbp]
        \centering
        \setlength{\abovecaptionskip}{0.cm}
        \includegraphics[width=0.8\textwidth]{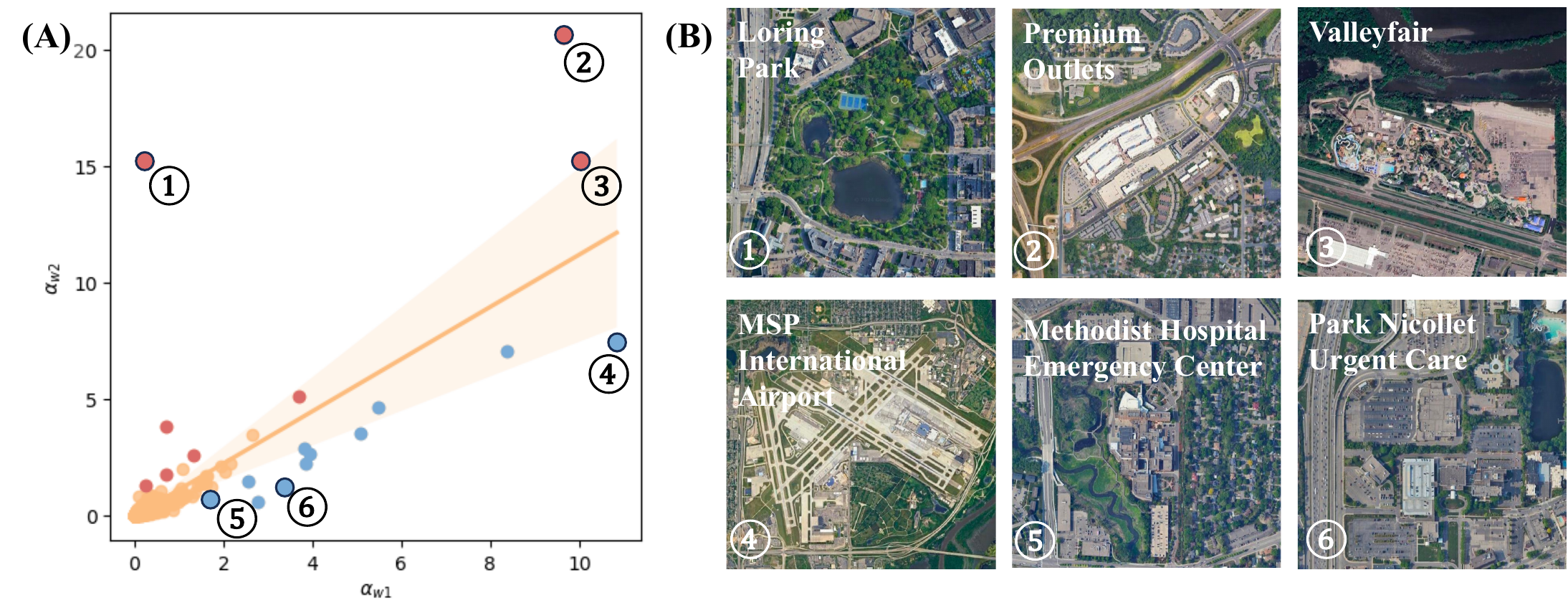}
        \caption{Lakeplace popularity disparity between weekdays and weekends. (A) presents the scatter plot of lakeplace human activity intensity on weekdays ($\alpha_{w1}$) and weekends ($\alpha_{w2}$). Two types of outliers are highlighted: red points representing weekend lakeplaces, and blue points representing weekday lakeplaces. (B) displays six lakeplaces belonging to the two types in (A), respectively.}
        \label{fig:07}
    \end{figure*}

Lakeplace popularity, reflected by average daily visitation, is only a part of the human's lake. To echo the general temporal patterns, we investigated the popularity disparity between weekdays and weekends. Similar to $\alpha_{d}$, popularity on weekdays and weekends are represented by $\alpha_{w1}$ and $\alpha_{w2}$, respectively. These two indicators are also based on daily visitations, but only count in visitations on weekdays and weekends, respectively. As indicated by Figure \ref{fig:07}A, the disparity is trivial for most lakeplaces. However, there are still several lakeplaces with significant imbalanced popularity between weekdays and weekends. Some lakeplaces are heavily visited during the weekends, while much more vacant during the weekdays, such as Loring Park (Figure \ref{fig:07}\ding{172}), Premium Outlets (Figure \ref{fig:07}\ding{173}) and Valleyfair (Figure \ref{fig:07}\ding{174}). On the contrary, there are lakeplaces with much higher popularity during the weekdays. Evidently, MSP International Airport turns out to be a typical ``weekday lakeplace'' (Figure \ref{fig:07}\ding{175}). Several health and medical facilities (e.g., Figure \ref{fig:07}B\ding{176} and Figure \ref{fig:07}B\ding{177}) are also being classified into this type. Note that all examples above are not lakeplaces with high-level on-lake contributions, as lakeplaces for tourism are usually not largely affected by working or not. When examining through the lens of daily popularity, the aforementioned lakeplaces are all among the most popular lakeplaces. The disparity between weekdays and weekends further depicts the popularity nuance, indicating that temporal scales are critical when sensing lakeplaces.

\subsection{Social-demographics for knowledge discovery}
\begin{figure}[H]
        \centering
        \setlength{\abovecaptionskip}{0.cm}
        \includegraphics[width=0.9\textwidth]{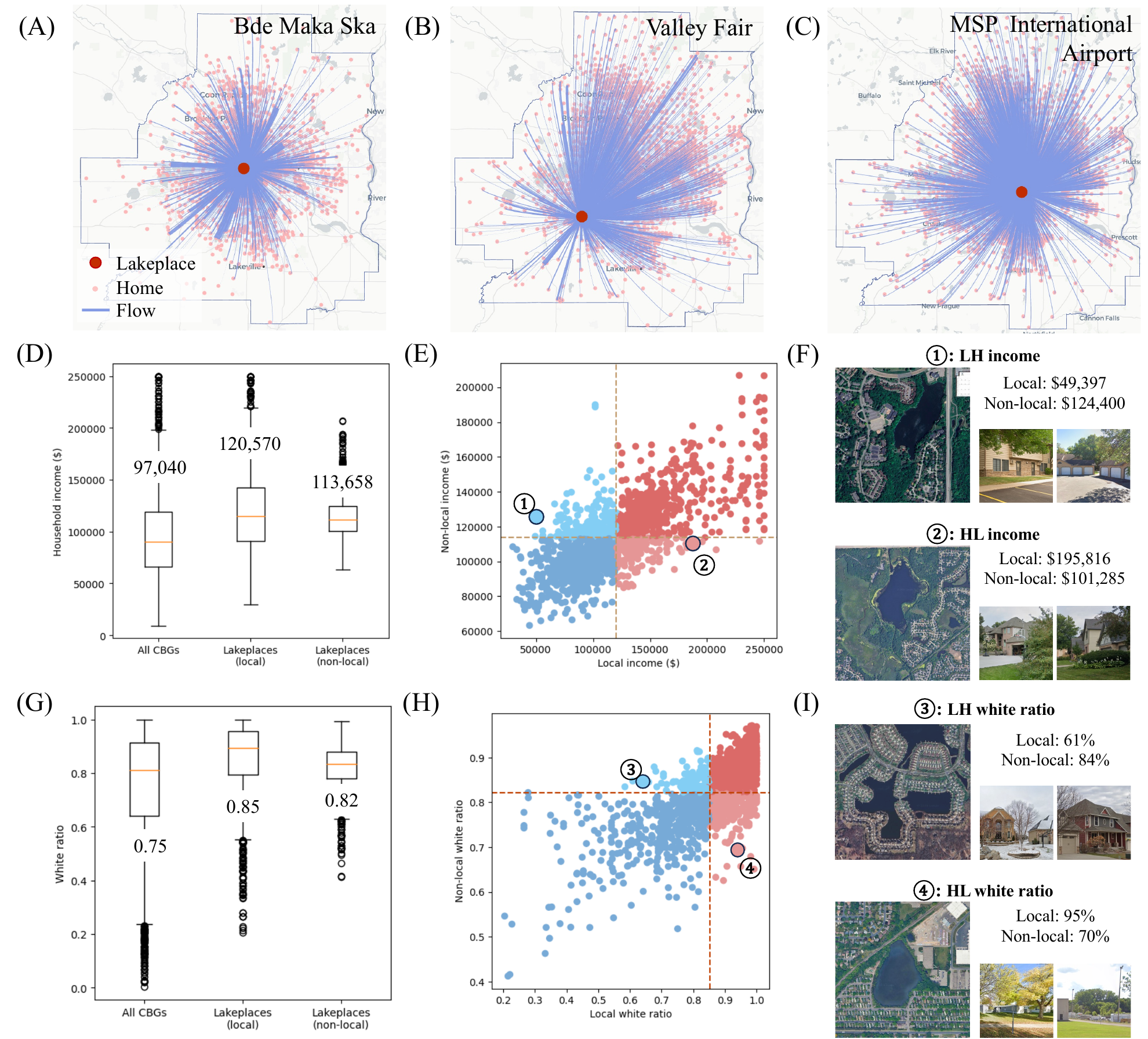}
        \caption{(A) to (C) display OD (Origin-Destination) flows from three different lakeplaces (Bde Maka Ska, Valley Fair, and MSP International Airport) to the residential areas of their visitors, respectively. For household income, (D) shows income disparity (on average) among all CBGs, locals and non-locals, (E) presents the income correlation between locals and non-locals, and (F) illustrates two typical examples of outliers: lakeplace \ding{172} and lakeplace \ding{173}. Similarly, (G) shows white ratio disparity (on average) among three groups in (D), (H) presents the white ratio of locals and non-locals for each lakeplace respectively, and (I) demonstrates two examples of outliers for white ratio: lakeplace \ding{174} and lakeplace \ding{175}.}
        \label{fig:08}
    \end{figure}

To further explore the essence of human's lakeplaces, we classified individuals who interact with a lakeplace into local (residents) and non-local (visitors) by identifying whether they live inside the visited lakeplace or not. For an individual, we inferred their social-demographic features based on their residential area and the corresponding aggregated census data. We chose the census block group (CBG), the smallest geographic unit containing social-demographic statistical information, to map social-demographic features from the geographic unit to each individual. In total, residential areas of 187,635 individuals were identified. For each individual, we extracted the longitude-latitude locations of all their visitations within the whole month, and applied the DBSCAN method to identify the most likely residential area \citep{huang2023reconstructing}. We created maps of Origin-Destination (OD) between three lakeplaces and the residential areas of their visitors based on home detection results, respectively (Figures \ref{fig:08}A-C). In terms of the spatial distribution of OD flows, the residential areas of visitors to Bde Maka Ska are more clustered, whereas those to MSP International Airport are more spatially dispersed. The disparity among the three OD flow patterns reflects differences in the attractiveness of lakeplaces, indicating potential social-demographic disparity of visitors to lakeplaces. 

We considered income and race as two key social-demographic features \citep{li2024behavior} for portraying visitors to lakeplaces, and probed into the disparity between residents and visitors for each lakeplace. As for income, the American Community Survey (ACS) median household income data for the 2021 year is applied to estimate individual income levels. Similarly, we applied race division data from the 2021 ACS to calculate the ratio of white people (white ratio) for each CBG. It is observed that people living inside or visiting lakeplaces are more likely to have higher income (Figure \ref{fig:08}D) and to be white people (Figure \ref{fig:08}G). Compared with non-locals, the average income level and white ratio are slightly higher for locals (both at the 1\% significance level), despite higher in-group variations. The comparison suggests that people who frequently interact with lakes are more likely to be wealthier and more white in the Twin Cities. It indicates lake-related environmental injustice, as higher accessibility to lakes may be a privilege predominantly shared by higher-income and white groups. One typical example is the lakeplace consisting of Cedar Lake (Figure \ref{fig:02}\ding{178}) and Lake of the Isles (Figure \ref{fig:02}\ding{179}), with the highest household income level and white ratio across the whole TCMA. In the 20th century, the area was once designated to be the settlement of white residents due to racial covenants, reflecting the preference of living near the water view by the privileged social groups \citep{walker2023making}.

As shown in Figure \ref{fig:08}E, there is a high correlation between the household income of locals and of non-locals. The same pattern is also shown in the white ratio correlation (Figure \ref{fig:08}H). Still, several meaningful outliers (Low-High (LH) or High-Low (HL)) are identified. Lakeplace \ding{172} (\ding{172} in Figure \ref{fig:08}F) is an example for LH income pattern, where local income is lower than that of non-local visitors, and lakeplace \ding{173} (\ding{173} in Figure \ref{fig:08}F) is a typical example of HL pattern. In comparison, lakeplace \ding{172} is surrounded by several apartments for low-income level groups, while lakeplace \ding{173} is located in the center of plenty of well-built private houses. Thus, the disparity of the built environment helps to explain the vast disparity between the local income of these two lakeplaces, as the non-local income level is similar to each other. Similarly, lakeplace \ding{174} (\ding{174} in Figure \ref{fig:08}I) and lakeplace \ding{175} (\ding{175} in Figure \ref{fig:08}I) present two opposite outliers of the white ratio pattern, respectively. The former one is a residential area preferred by the black group and surrounded by plenty of old-fashioned houses, and the latter one is a commercial area occupied by low-income white residents and colored customers. Based on the metric of popularity, these four lakeplaces may be neglected or be regarded as``anonymous'', but they come alive and stand out in lakeplace social-demographic profiling.


\section{Lakeplace sensing framework}\label{framework}

The practical way towards human's lakes can be summarized as three steps, including (1) human-environment information integration, (2) spatiotemporal analysis, and (3) knowledge discovery, as illustrated in Figure \ref{fig:09}. Through the processes, we aim to reflect interactions between humans and lakes based on human digital traces, investigate the spatiotemporal characteristics of human activities on and around the lakes, and discover social-demographic knowledge to profile lakes. The framework echoes the theoretical discussions on transforming lakes into lakeplaces (Figure \ref{fig:01}), providing a way to render meaningful lakes by treating them as unique places. Details of the lakeplace sensing framework are described below:

\begin{figure*}[htbp!]
        \centering
        \setlength{\abovecaptionskip}{0.cm}
        \includegraphics[width=\textwidth]{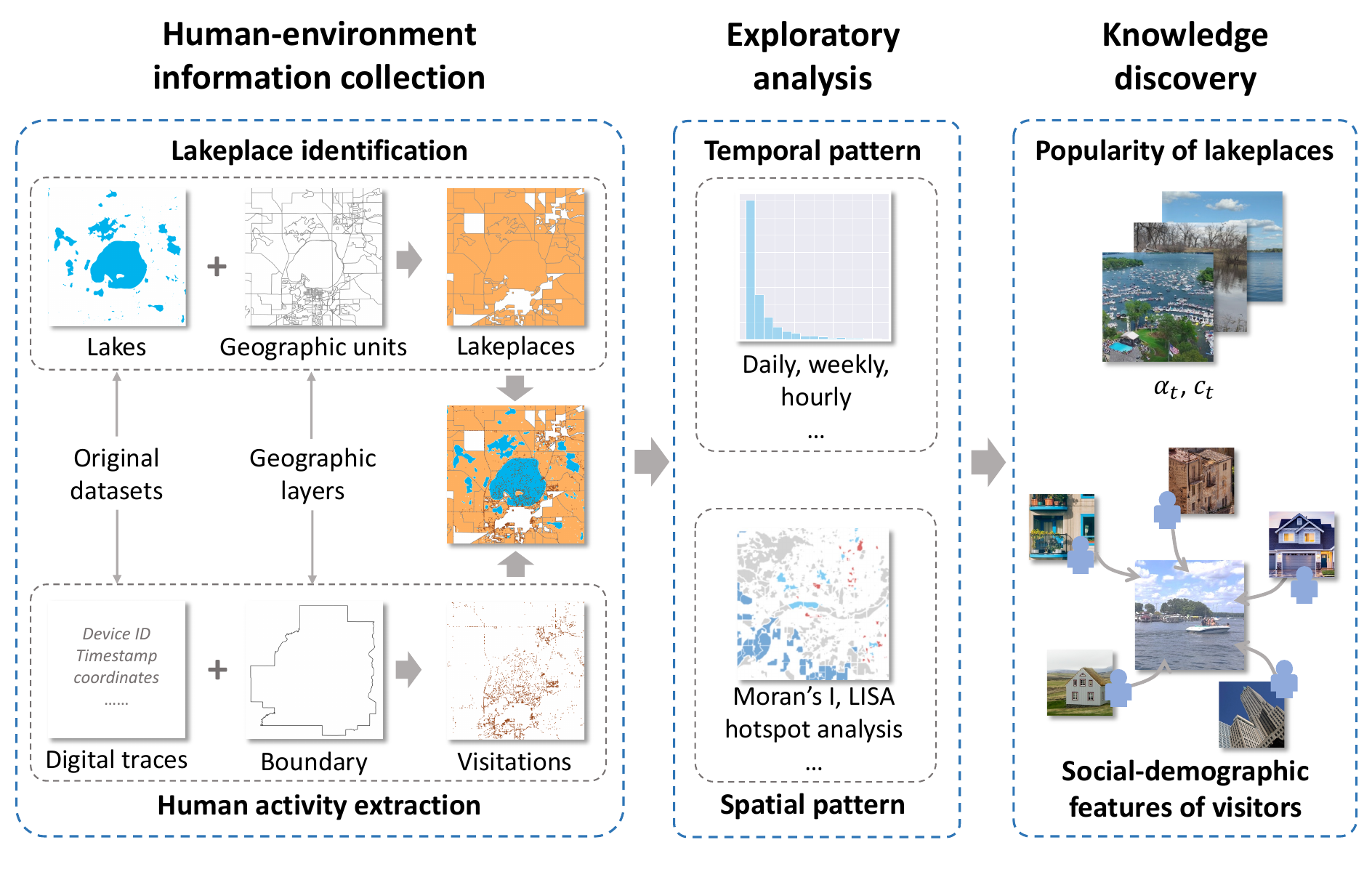}
        \caption{Lakeplace sensing framework.}
        \label{fig:09}
    \end{figure*}

\textit{Step 1: Human-environment information collection}

The first step is to project human activities on corresponding locations, serving as the basis of this framework and containing two parts: lakeplace identification and human activity extraction. To identify lakeplaces in bulk, several rules are set based on spatial relationships between lakes and nearby geographic units. For simplicity, all first-order contiguous units are merged to generate the lakeplace for each lake by spatial selection and aggregation. Intersected lakeplaces are merged to ensure that they don't overlap with each other. In this way, a lake is encompassed by its lakeplace, and there may be one or more lakes inside one lakeplace. Lake-related human activities can be extracted from large digital-trace datasets by projecting visitations on the lakeplace layer referring to their coordinates. Once human activity information and the areas of lakeplaces are collected, visitation records are mapped on the lakeplace layer for the following lake-human interaction analyses.

\textit{Step 2: Spatiotemporal analysis}

The second step is designed for discovering temporal variation patterns as an in-between step to map spatial and temporal characteristics of human activities related to lakeplaces. Temporal pattern mainly depicts human activity intensity inside lakeplaces at different temporal scales, such as visitations on each day and each hour. Clustering methods, such as hierarchical clustering, may be applied to identify different temporal patterns of lakeplaces based on hourly visitation changes and daily visitation changes, respectively \citep{zhu2017street}. Similarly, spatial pattern analysis is mainly about showing the spatial distribution and clustering characteristics of human activities through exploratory data analysis (EDA) methods. For instance, Moran's I can be employed to detect spatial auto-correlation of the visitations of lakeplaces, indicating meaningful spatial clusters based on human activity intensity.
    
\textit{Step 3: Knowledge discovery}

The third step is based on all the analyses in the aforementioned two sections. To answer the fundamental question: ``\textit{which are the most popular lakes?}'', average visitations inside lakeplaces at different temporal scales are scaled and ranked to represent the popularity of lakeplaces. Bearing that the total amount of visitations commonly increases by the area of lakeplace, an area-based scalar $\alpha_t$ is applied to represent human activity intensity at a certain temporal scale $t$ for each lakeplace:

\begin{equation}
    \alpha_t = \frac{vp_t}{ap} * \frac{1}{\sqrt{rank(vp_t)}},
\label{Equation 1}
\end{equation}

\noindent
where $vp_t$ denotes the average number of visitations on the lakeplace at temporal scale $t$, such as several days or several hours, and $ap$ refers to the covering area of this lakeplace. Instead of z-score, min-max normalization, or other standardization methods, we chose to employ a ranking method to scale human activity intensities of lakeplaces since the geographic data follows the power distribution in most cases. For any lakeplace, $rank(vp_t)$ represents the ranking of its average lakeplace visitations at temporal scale $t$. The lower the value of $rank(vp_t)$, the higher the human activity intensity. For example, $\alpha_{d}$ refers to the daily human activity intensity based on the scaled daily average visitations (Section \ref{sec:On-lake}), and $\alpha_{w1}$ represents the weekday daily human activity intensity based on the scaled daily average visitations on weekdays (Section \ref{sec:Comparing}). They are used to measure lakeplace popularity at different temporal scales in our lakeplace sensing practice in the TCMA.

Comparing the number of visitations to lakes and corresponding lakeplaces helps to answer another important question: ``\textit{why are these lakeplace popular?}''. In other words, is it due to the lake or non-lake areas? For instance, a lake famous for on-lake recreation like fishing, swimming, and boating is likely to belong to the former type, while a small artificial lake inside a commercial center is a typical example of the latter type, as people visit that area for shopping or dining, not to interact with the lake. Here, to represent the proportion of on-lake visitations among all visitations inside the corresponding lakeplace at temporal scale $t$, we proposed another index, namely on-lake contribution $c_t$:

\begin{equation}
 c_t = \frac{vl_t} {vp_t}, 
\label{Equation 2}
\end{equation}

\noindent
where $vl_t$ and $vp_t$ indicate the number of on-lake visitations and of visitations inside the corresponding lakeplace at temporal scale $t$, respectively. The higher the value of $c_t$, the more human activities are brought by on-lake visitations as well as more intense interaction between humans and the lake. $c_d$, for instance, refers to the number of daily average on-lake visitations and of daily average visitations, which is applied to explain the popular lakeplaces by $\alpha_{d}$ in the TCMA (Section \ref{sec:On-lake}).

Furthermore, apart from lake popularity based on human activity intensity at multiple temporal scales, lakes are profiled based on the social-demographic features of their visitors. It is accomplished by identifying visitors' residential areas, extracting their social-demographic characteristics, and merging results for each lake. Two commonly used indicators, income and race, are employed to tell us what the visitors are like. Future lakeplace sensing may require the integration of multi-source and multi-modal datasets to discover more knowledge of lakeplaces, such as street image data and survey data \citep{jing2023exploring}. They can be utilized to illustrate the surroundings of lakeplaces more comprehensively and quantitatively, and offer individual social-demographic information. Moreover, the current way of sensing lakeplaces is objective, based on visitation data and general indicators. Semantic analysis may contribute to uncovering subjective lakeplaces based on human experience and emotion attached to lakeplaces.

\section{Discussion: Scale and more}
Despite efforts to construct and interpret lakeplaces, uncertainties still exist from the quantitative perspective due to the geographic scale variations. Lakeplace defined and identified in this paper, specifically, should be termed as ``census-block based lakeplace''. Most critically, the Modifiable Areal Unit Problem (MAUP) exists when generating lakeplaces \citep{wong2004modifiable}. The rationale behind our selection for the case study is empirical, bearing that the census block is the smallest administrative geographic unit and reflects neighborhood-scale human daily activities in the US. In other words, to what spatial extent lakes engage in human daily life has been predefined, as it is related to the intensity people interact with lakes. Boundaries of lakeplaces and human-lake interactions can be redefined and reformulated as spatial scale changes. As such, other sets of lakeplace layers based on geographic units, smaller or larger than the census block, may be required to examine the uncertainties and geographic context embedded in the concept of lakeplace \citep{kwan2012uncertain}.

The second facet of scale heads towards a more practical perspective. Is there any possibility for lakeplace, compared to the authorized and accepted geographic units such as census blocks, block groups, and tracts, to become a common geographic unit for urban studies? Indeed, apart from great potential as a novel perspective sensing lakes, lakeplace might be an ideal complement to statistics by static census survey. It is envisioned that the lakeplace unit system provides multidimensional census information at more flexible spatial and temporal scales, contributing to data fusion in urban database \citep{liu2020urban}. For instance, in addition to residents of lakeplaces, human activities along with social-demographic features of visitors to lakeplaces can also be captured. The selection of visitors can vary across different temporal scales and visit frequencies, depending on the research needs, such as focusing on frequent visitors within a week or all visitors within a month. Despite the appealing prospects, two issues remain to be reconsidered. One thing lies in the origin of lakeplace, which is yet based on an existing zoning scheme. That said, the concept of lakeplace has not transcended static and exact geographic boundaries. It still falls short in fully embracing data-driven approaches. Exploring unfixed boundaries of lakeplaces, such as buffering lakes by different accessibility indexes, may emerge as a new avenue for research and practice in urban planning. Another thing that lakeplaces differ from most geographic units is spatial coverage, as territories far from lakes are left uncharted. The zoning is inconsistent in space, more like a thematic map of lakes rather than the so-called geographic unit system. It calls for further research to establish connections between the lakeplace layer and traditional demographic maps covering the entire area, such as population, income, and racial maps. Nevertheless, lakeplace as a geographic unit yields important insights into urban planning and management, entailing more embodied conceptualization and broader enrichment via future efforts.

Another aspect of the scale problem is the distance. In this study, a simple dichotomy in identifying local and non-local visitors for each lakeplace leads to some debates on the categorization. When defining the local visitors by whether the individual's residential area is inside the lakeplace, the underlying logic inevitably causes a couple of questions: is it rational to accept this dichotomy, and how far is considered the scope of home? By way of example, a person may reside slightly outside the exact boundary of lakeplace, but their routine still centers on the corresponding lake, serving as an anchor point for their daily life. The boundary thus ought to be fuzzy in daily life and human experience attached to the lake. Conceptual disputes are not the only thing rooted in the distance question, as human movement data applied in our study is also packed with uncertainties. Location accuracy and sampling could probably affect a series of judgments, such as home location identification, as well as whether people are on the lakes. For instance, jogging around the lake and boating on the lake may take place within only 5 meters, which may be recorded as both on the lake. Similarly, there will be bias when measuring the popularity of lakeplaces with highway crossings or near highways. As such, it raises another question for us: what specific activity type does the record represent? That's what human movement data is unable to resolve due to a lack of semantic information. We have grabbed the knowledge about built environment data to infer human activity types, but still at a relatively qualitative and collective level.

Lakeplace entails more efforts in the future. Work should be directed toward advancing knowledge discovery and quantification in practice to benefit quality of life. There are diverse research topics with great promise, for instance, prediction of lakeplace utilization, evaluation of lakeplace socioeconomic, and modeling the optimal scale of lakeplace. More importantly, being the first study of lakeplace, we call for more exploration from the qualitative field to expand the horizon of lakeplace studies. We envision lakeplace as an avenue in which researchers and decision-makers rethink the values of urban freshwater systems, understand human-environment interactions, and collaborate to promote urban sustainability.

\bibliographystyle{elsarticle-harv} 
\bibliography{elsarticle-template-harv}

\end{document}